\documentstyle[12pt,aasms4]{article}

\begin{document}

\title{Bias and Hierarchical Clustering}

\author{Peter Coles\altaffilmark{1,2},
        Adrian L. Melott\altaffilmark{3},
         Dipak Munshi\altaffilmark{2,4,5}}

\altaffiltext{1}{School of Physics \& Astronomy, University of
Nottingham, University Park, Nottingham NG7 2RD, UK}

\altaffiltext{2}{Astronomy Unit, Queen Mary \& Westfield College,
University of London, London E1 4NS, UK}

\altaffiltext{3}{Department of Physics and Astronomy, University
of Kansas, Lawrence, KS  66045, USA}

\altaffiltext{4}{International School for Advanced Studies
(SISSA), Via Beirut 2-4, I-34013 Trieste, Italy}

\altaffiltext{5}{Max Planck Institut fur Astrophysik, Karl
Schwarzschild Str. 1, Postfach 1523, D-85740 Garching, Germany}

\begin{abstract}
It is now well established that galaxies are biased tracers of the
distribution of matter, although it is still not known what form
this bias takes. In local bias models  the propensity for a galaxy
to form at a point depends only on the overall density of matter
at that point. Hierarchical scaling arguments allow one to build a
fully-specified model of the underlying distribution of matter and
to explore the effects of local bias in the regime of strong
clustering. Using a generating-function method developed by
Bernardeau \& Schaeffer (1992), we show that hierarchical models
lead one directly to the conclusion that a local bias does not
alter the shape of the galaxy correlation function relative to the
matter correlation function on large scales. This provides an
elegant extension of a result first obtained by Coles (1993) for
Gaussian underlying fields and confirms the conclusions of
Scherrer \& Weinberg (1998) obtained using a different approach.
We also argue that particularly dense regions in a hierarchical
density field display a form of bias that is different from that
obtained by selecting such peaks in Gaussian fields: they are
themselves hierarchically distributed with scaling parameters
$S_p=p^{(p-2)}$. This kind of bias is also factorizable, thus in
principle furnishing a simple test of this class of models.
\end{abstract}

\keywords{galaxies: statistics -- large scale structure of the
Universe}

\section{Introduction}
The biggest stumbling-block for attempts to confront theories of
cosmological structure formation with observations of galaxy
clustering is the uncertain and possibly biased relationship
between galaxies and the distribution of gravitating matter. The
idea that galaxy formation might be biased goes back to the
realization by Kaiser (1984) that the reason Abell clusters
display stronger correlations than galaxies at a given separation
is that these objects are selected to be particularly dense
concentrations of matter. As such, they are very rare events,
occurring in the tail of the distribution function of density
fluctuations. Under such conditions a ``high-peak'' bias prevails:
rare high peaks are much more strongly clustered than more typical
fluctuations (Bardeen et al. 1986). If the properties of a galaxy
(its morphology, color, luminosity) are influenced by the density
of its parent halo, for example, then differently-selected
galaxies are expected to a different bias (e.g. Dekel \& Rees
1987). Observations show that different kinds of galaxy do cluster
in different ways (e.g. Loveday et al. 1995; Hermit et al. 1996).

In {\em local bias} models, the propensity of a galaxy to form at
a point where the total (local) density of matter is $\rho$ is
taken to be some function $f(\rho)$ (Coles 1993, hereafter C93;
Fry \& Gaztanaga 1993, hereafter FG93). It is possible to place
stringent constraints on the effect this kind of bias can have on
galaxy clustering statistics without making any particular
assumption about the form of $f$. In this {\it Letter}, we
describe  the results of a different approach to local bias models
that exploits new results from the theory of hierarchical
clustering in order to place stronger constraints on what a local
bias can do to galaxy clustering. We leave the technical details
to  Munshi et al. (1999a,b) and Bernardeau \& Schaeffer (1999);
here we shall simply motivate and present the results and explain
their importance in a wider context.

\section{Hierarchical Clustering}
The fact that Newtonian gravity is scale-free suggests that the
$N$--point correlation functions of self-gravitating particles,
$\xi_N$, evolved into the large-fluctuation regime by the action
of gravity, should obey a scaling relation of the form
\begin{equation}
\xi_p( \lambda {\bf r}_1, \dots  \lambda {\bf r}_p ) =
\lambda^{-\gamma(p-1)} \xi_p( {\bf r}_1, \dots {\bf r}_p )
\label{hierarchical}
\end{equation}
when the elements of a structure are scaled by a factor $\lambda$
(e.g. Balian \& Schaeffer 1989). Observations offer some support
for such an idea, in that the observed two-point correlation
function $\xi(r)$ of galaxies is reasonably well represented by a
power law over a large range of length scales,
\begin{equation} \xi({\bf r}) = \Big ( {r \over 5h^{-1} {\rm Mpc}}
\Big )^{-1.8} \end{equation} (Groth \& Peebles 1977; Davis \& Peebles
1977) for $r$ between, say, $100 h^{-1} {\rm kpc}$ and $
10h^{-1}~{\rm~ Mpc}$. The observed three point function, $\xi_3$, is
well-established to have a hierarchical form \begin{equation}
\xi_3({\bf x}_a, {\bf x}_b, {\bf x}_c) = Q[\xi_{ab}\xi_{bc} +
\xi_{ac}\xi_{ab} + \xi_{ac}\xi_{bc}], \end{equation} where
$\xi_{ab}=\xi ({\bf x}_a, {\bf x}_b)$, etc, and $Q$ is a constant
(Davis \& Peebles 1977; Groth \& Peebles 1977). The four-point
correlation function can be expressed as a combination of  graphs with
two different topologies -- ``snake'' and ``star'' -- with
corresponding (constant) amplitudes $R_a$ and $R_b$ respectively:
\begin{equation} \xi_4({\bf x}_a, {\bf x}_b, {\bf x}_c, {\bf x}_d) =
R_a[\xi_{ab}\xi_{bc}\xi_{cd} + \dots ({\rm 12~terms})] +
R_b[\xi_{ab}\xi_{ac}\xi_{ad} + \dots ({\rm 4~terms})] \end{equation}
(e.g. Fry \& Peebles 1978; Fry 1984).

It is natural to guess that all p-point correlation functions can be
expressed as a sum over all possible p-tree graphs with (in general)
different amplitudes $Q_{p,\alpha}$  for each tree diagram topology
$\alpha$. If it is further assumed that there is no  dependence of
these amplitudes upon the shape of the diagram, rather than its
topology, the correlation functions should obey the following
relation: \begin{equation} \xi_p( {\bf r}_1, \dots {\bf r}_p ) =
\sum_{\alpha, ~p-{\rm trees}} Q_{p,\alpha} \sum_{\rm labellings}
\prod_{\rm edges}^{(p-1)} \xi({\bf r}_i, {\bf r}_j). \end{equation} To
go further it is necessary to find a way of calculating $Q_p$. One
possibility, which appears remarkably successful when compared with
numerical experiments (Munshi et al. 1999b; Bernardeau \& Schaeffer
1999), is to calculate the amplitude for a given graph by simply
assigning a weight to each vertex of the diagram $\nu_n$, where $n$ is
the order of the vertex (the number of lines that come out of it), 
regardless of the topology of the diagram in which it occurs. In this case
\begin{equation} Q_{p,\alpha}=\prod_{\rm vertices} \nu_n.
\end{equation} Averages of higher-order correlation functions can be
defined as \begin{equation} \bar{\xi}_p=\frac{1}{V^p} \int \ldots \int
\xi_p({\bf r}_1\ldots {\bf r_p}) dV_1\ldots dV_p. \label{eq:xibar}
\end{equation} Higher-order statistical properties of galaxy counts
are often described in terms of the scaling parameters $S_p$
constructed from the $\bar{\xi}_p$ via \begin{equation}
S_p=\frac{\bar{\xi}_p}{\bar{\xi}_2^{p-1}}.\label{eq:s_p}
\end{equation} It is a consequence of the particular class of hierarchical clustering models
defined by equations (5) \& (6) that {\it all} the $S_p$ should be
constant, independent of scale.

\section{Local Bias}

Using a generating function technique, originally developed by
Bernardeau \& Schaeffer (1992), it is possible to derive a series
expansion for the $m$-point count probability distribution
function of the objects $P_m(N_1,....N_m)$ (the joint probability
of finding $N_i$ objects in the $i$-th cell, where $i$ runs from $1$ to $m$) 
from the $\nu_n$. The
hierarchical model outlined above is therefore statistically
complete. In principle, therefore, any statistical property of the
evolved distribution of matter can be calculated just as it can
for a Gaussian random field. This allows us to extend various
results concerning the effects of biasing on the initial
conditions into the nonlinear regime in a more elegant way than is
possible using other approaches to hierarchical clustering.

For example, let us consider the joint occupation probability
$P_2(N_1, N_2)$ for two cells to contain $N_1$ and $N_2$ particles
respectively. Using the generating-function approach outlined
above, it is quite easy to show that, at lowest order,
\begin{equation}
P_2(N_1,N_2)=P_1(N_1)P_1(N_2)+P_1(N_1)b(N_1)P_1(N_2)b(N_2)\xi_{12}(r_{12
}), \label{eq:2pt} \end{equation}
 where the $P_1(N_i)$ are the
individual count probabilities of each volume separately and
$\xi_{12}$ is the underlying mass correlation function. The
function $b(N_i)$ we have introduced in (9) depends on the set of
$\nu_n$ appearing in equation (6); its precise form does not
matter in this context, but the structure of equation (9) is very
useful. We can use (9) to define 
\begin{equation}
1+\xi_{N_1 N_2}(r_{12}) \equiv \frac{P(N_1, N_2)}{P_1(N_1)P_1(N_2)},
\end{equation}
where $\xi_{N_1 N_2}(r_{12})$ is the cross-correlation of
``cells'' of occupancy $N_1$ and $N_2$ respectively. From this definition
and equation (9) it follows that
\begin{equation} \xi_{N_1 N_2}(r)=b(N_1)b(N_2)\xi_{12}(r);
\label{eq:2bias}
\end{equation} 
we have dropped the subscripts on $r$ for clarity from now on.
>From (11) we can obtain 
\begin{equation}
b_{N}^2(r_{12}) = \frac{\xi_{N N}(r)}{\xi_{12}(r)}
\end{equation}
for the special case where $N_1=N_2=N$ which can be identified with the usual
definition of the bias parameter associated with the correlations among a 
given set of objects $\xi_{\rm obj}(r)=b^{2}_{\rm obj} \xi_{\rm mass}(r)$.
Moreover, note that at this order (which is valid on large
scales), the correlation bias defined by equation (11)
factorizes into contributions $b_{N_i}$ from each individual cell (Bernardeau 1996; 
Munshi et al. 1999b).

Coles (1993) proved, under weak conditions on the form of a local
bias $f(\rho)$ as discussed in the introduction, that the
large-scale biased correlation function would generally have a
leading order term proportional to $\xi_{12}(r_{12})$. In other
words, one cannot change the large-scale slope of the correlation
function of locally-biased galaxies with respect to that of the
mass. This ``theorem'' was proved for bias applied to Gaussian
fluctuations only and therefore does not obviously apply to galaxy
clustering, since even on large scales deviations from Gaussian
behaviour are significant. It also has a more minor loophole,
which is that for certain peculiar forms of $f$ the leading order
term is proportional to $\xi_{12}^2$, which falls off more sharply
than $\xi_{12}$ on large scales.

Steps towards the plugging of this gap began with FG93 who used an
expansion of $f$ in powers of $\delta$ and weakly non-linear
(perturbative) calculations of $\xi_{12}(r)$ to explore the
statistical consequences of biasing in more realistic (i.e.
non-Gaussian) fields. Based largely on these  arguments, Scherrer
\& Weinberg (1998), hereafter SW98, confirmed the validity of the C93
result in the non-linear regime, and also showed explicitly that
non-linear evolution always guarantees the existence of a linear
leading-order term regardless of $f$, thus plugging the small gap
in the original C93 argument. These works have
a similar motivation to ours, and also exploit hierarchical
scaling arguments of the type discussed in \S 2 {\em en route} to
their conclusions. What is different about the approach we have
used in this paper is that the somewhat cumbersome simultaneous 
expansion of $f$
and $\xi_{12}$ used by SW98 is not required in our calculation: we use the
generating functions to proceed directly to the joint probability
(9), while SW98 have to perform a complicated sum over moments of a
bivariate distribution. The factorization of the probability
distribution (9) is also a stronger result than that presented by SW98,
in that it leads almost trivially to the C93 ``theorem'' but also
generalizes to higher-order correlations than the two-point case
under discussion here.

Note that the density of a cell of given volume is simply
proportional to its occupation number $N$. The factorizability of
the dependence of $\xi_{N_1 N_2}(r_{12})$ upon $b(N_1)$ and
$b(N_2)$ in (11) means that applying a local bias $f(\rho)$ boils
down to applying some bias function $F(N)=f[b(N)]$ to each cell.
Integrating over all $N$ thus leads directly to the same
conclusion as C93, i.e. that the large-scale $\xi(r)$ of
locally-biased objects is proportional to the underlying matter
correlation function. This has also been confirmed by numerically
using $N$-body experiments (Mann et al. 1998; Narayanan et al.
1998).

\section{Halo Bias}

In hierarchical models, galaxy formation involves the following
three stages:
\begin{enumerate}
\item the formation of a dark matter halo;
\item the settling of gas into the halo potential;
\item the cooling and fragmentation of this gas into stars.
\end{enumerate}
Rather than attempting to model these stages in one go by a simple
function $f$ of the underlying density field it is interesting to see
how each of these selections might influence the resulting statistical
properties. Bardeen et al. (1986), inspired by Kaiser (1984),
pioneered this approach by calculating detailed statistical properties
of high-density regions in Gaussian fluctuations fields. Mo \& White
(1996) and Mo et al. (1997) went further along this road by using an
extension of the Press-Schechter (1974) theory to calculate the
correlation bias of halos, thus making an attempt to correct for the
dynamical evolution absent in the Bardeen et al. approach. The
extended Press-Schechter approach seem to be in good agreement with
numerical simulations, except for small halo masses (Jing 1998). It
forms the basis of many models for halo bias in the subsequent
literature (e.g. Moscardini et al. 1998; Tegmark \& Peebles 1998).

The hierarchical models furnish an elegant extension of this work that
incorporates both density-selection and non-linear dynamics in an
alternative to the  Mo \& White (1996) approach. We  exploit the properties of
equation (\ref{eq:2pt}) to construct the correlation function of
volumes where the occupation number exceeds some critical value. For
very high occupations these volumes should be in good correspondence
with collapsed objects.

The way of proceeding is to construct a tree graph for all the
points in both volumes. One then has to re-partition the elements of
this  graph into internal lines (representing the correlations within
each cell) and external lines (representing inter-cell correlations).
Using this approach the distribution of high-density regions in a
field whose correlations are given by eq. (5) can be shown to be
itself described by a hierarchical model, but one in which the vertex
weights, say $M_n$, are different from the underlying weights $\nu_n$
(Bernardeau \& Schaeffer 1992, 1999; Munshi et al. 1999a,b).

First note that a density threshold is in fact a form of local
bias, so the effects of halo bias are governed by the same
strictures as described in the previous section. Many of the other
statistical properties of the distribution of dense regions can be
reduced to a dependence on a scaling parameter $x$, where
\begin{equation} x=N/N_c. \end{equation} In this definition
$N_c=\bar{N}\bar{\xi}_2$, where $\bar{N}$ is the mean number of
objects in the cell and $\bar{\xi}_2$ is defined by eq.
(\ref{eq:xibar}) with $p=2$.  The scaling parameters $S_p$  can be
calculated as functions of $x$, but are generally rather messy (Munshi
et al. 1999a). The most interesting limit when $x\gg 1$ is, however,
rather simple. This is because the vertex weights describing the
distribution of halos depend only on the $\nu_n$ and this dependence
cancels in the ratio (\ref{eq:s_p}). In this regime, \begin{equation}
S_p=p^{(p-2)} \end{equation} for all possible hierarchical models. The
reader is referred to Munshi et al. (1999a) for details. This result
is also obtained in the corresponding limit for very massive halos by
Mo et al. (1997). The agreement between these two very different
calculations supports the inference that this is a robust prediction
for the bias inherent in dense regions of a distribution of objects
undergoing gravity-driven hierarchical clustering.

\section{Discussion and Conclusions}

The main purpose of this {\em Letter} has been to advertise the
importance of recent developments in the theory of
gravitational-driven hierarchical clustering. The model described in
equations (5) \& (6) provides a statistically-complete prescription
for a density field that has undergone hierarchical clustering. This
allows us to improve considerably upon biasing arguments based on an
underlying Gaussian field.

These methods allow a simpler proof of the result obtained by SW98
that  strong non-linear evolution does not invalidate the local
bias theorem of C93. They also imply that the effect of bias on a
hierarchical density field is factorizable. A special case of this
is the bias induced by selecting regions above a density
threshold. The separability of bias predicted in this kind of
model could be put to the test if a population of objects could be
found whose observed characteristics (luminosity, morphology,
etc.) were known to be in one-to-one correspondence with the halo
mass. Likewise, the generic prediction of higher-order correlation
behaviour described by the behaviour of $S_p$ in equation
(\ref{eq:s_p}) can also be used to construct a test of this
particular form of bias.

Referring to the three stages of galaxy formation described in \S 4,
analytic theory has now developed to the point where it is fairly
convincing on (1) the formation of halos. Numerical experiments are
beginning now to handle (2) the behaviour of the gas component
(Blanton et al. 1998, 1999). But it is unlikely that much will be
learned about (3) by theoretical arguments in the near future as the
physics involved is poorly understood (though see Benson et al. 1999).
Arguments have already been advanced to suggest that bias might not be
a deterministic function of $\rho$, perhaps because of stochastic or
other hidden effects (Dekel \& Lahav 1998; Tegmark \& Bromley 1999).
It also remains possible that large-scale non-local bias might be
induced by environmental effects (Babul \& White 1991; Bower et al.
1993).

Before adopting these more complex models, however, it is
important to exclude the simplest ones, or at least deal with that
part of the bias that is attributable to known physics. At this stage
this means that the `minimal' bias model should be that based on the
selection of dark matter halos. Establishing the extent to which
observed galaxy biases can be explained in this minimal way is clearly
an important task.

\acknowledgements

ALM acknowledges the support of the NSF-EPSCoR program, and DM
acknowledges the receipt of an Alexander von Humboldt research
fellowship. We thank Bob Scherrer for interesting discussions;
Francis Bernardaeau and Richard Schaeffer for helpful comments; and the
referee, David Weinberg, for extremely helpful criticisms of an earlier
version of this paper.


\begin{thebibliography}{}

\bibitem[Babul \& White 1991]{bw91}
Babul, A., \& White, S.D.M., 1991, MNRAS, 253, L31

\bibitem[Bardeen et al. 1986]{bbks}
Bardeen, J.M., Bond, J.R., Kaiser, N., \& Szalay, A.S., 1986, ApJ,
304, 15

\bibitem[Benson et al. 1999]{be99}
Benson, A.J., Cole, S., Frenk, C.S., Baugh, C.M., \& Lacey, C.G.,
1999, MNRAS, submitted, astro-ph/9903343

\bibitem[Bernardeau 1996]{b96}
Bernardeau, F., 1996, A\& A, 312, 11

\bibitem[Bernardeau \& Schaeffer 1992]{bs92}
Bernardeau, F., \& Schaeffer, R., 1992, A\& A, 255, 1

\bibitem[Bernardeau \& Schaeffer 1999]{bs99}
Bernardeau, F., \& Schaeffer, R., 1999, A\& A, submitted,
astro-ph/9903387

\bibitem[Blanton et al. 1998]{bl98}
Blanton, M., Cen, R., Ostriker, J.P., \& Strauss, M.A., 1998, ApJ,
submitted, astro-ph/9807029

\bibitem[Blanton et al. 1998]{bl99}
Blanton, M., Cen, R., Ostriker, J.P., Strauss, M.A., \& Tegmark, M.,
1999, ApJ, submitted, astro-ph/9903165


\bibitem[Bower et al. 1993]{b93}
Bower, R.G., Coles, P., Frenk, C.S., \& White, S.D.M. 1993, ApJ,
405, 403

\bibitem[Coles 1993]{c93}
Coles, P., 1993, MNRAS, 262, 1065 (C93)

\bibitem[Davis \& Peebles 1977]{dp77}
Davis, M., \& Peebles, P.J.E., 1977, ApJS, 34, 425

\bibitem[Dekel \& Lahav 1998]{dl98}
Dekel, A., \& Lahav, O., 1998, preprint, astro-ph/9806193

\bibitem[Dekel \& Rees 1987]{dr87}
Dekel, A., \& Rees, M.J., 1987, Nature, 326, 455

\bibitem[Fry 1984]{f84}
Fry, J.N., 1984, ApJ, 279, 499

\bibitem[Fry \& Gaztanaga 1993]{fg93}
Fry, J.N., \& Gaztanaga, E., 1993, ApJ, 413, 447 (FG93)

\bibitem[Fry \& Peebles 1978]{fp78}
Fry, J.N., \& Peebles, P.J.E., 1978, ApJ, 221, 19

\bibitem[Groth \& Peebles 1977]{gp77}
Groth, E., \& Peebles, P.J.E., 1977, ApJ, 217, 385

\bibitem[Hermit et al. 1996]{h96}
Hermit, S., Santiago, B.X., Lahav, O., Strauss, M.A., Davis, M.,
Dressler, A., \& Huchra, J.P., 1996, MNRAS, 283, 709


\bibitem[Kaiser 1984]{k84} Kaiser, N., 1984, ApJ, 284, L9

\bibitem[Loveday et al. 1995]{love}
Loveday, J., Maddox, S.J., Efstathiou, G., Peterson, B.A., 1995,
ApJ, 442, 557

\bibitem[Mann et al. 1998]{mann}
Mann, R.G., Peacock, J.A., \& Heavens, A.F., 1998, MNRAS, 293, 209

\bibitem[Mo \& White 1996]{mw96}
Mo, H.-J., \& White, S.D.M., 1996, MNRAS, 282, 347

\bibitem[Mo, Jing \& White 1997]{mjw97}
Mo, H.-J., Jing, Y.-P., \& White, S.D.M., 1997, MNRAS, 284, 189

\bibitem[Moscardini et al. 1998]{mosc}
Moscardini, L., Coles, P., Lucchin, F., \& Matarrese, S., 1998,
MNRAS, 299, 95

\bibitem[Munshi et al. 1999a]{m99a}
Munshi, D., Coles, P., \& Melott, A.L., 1999a, MNRAS, in press,
astro-ph/9812337

\bibitem[Munshi et al. 1999b]{m99b}
Munshi, D., Coles, P., \& Melott, A.L., 1999b, MNRAS, in press,
astro-ph/9902215

\bibitem[Narayanan et al. 1999]{n99}
Narayanan, V.K., Berlind, A.A., \& Weinberg, D.H., 1998, preprint,
astro-ph/9812002

\bibitem[Press \& Schechter 1974]{ps74}
Press, W.H. \& Schechter, P.L., 1974, ApJ, 187, 425

\bibitem[Scherrer \& Weinberg 1998]{sw98}
Scherrer, R.J., \& Weinberg, D.H., 1998, ApJ, 504, 607 (SW98)

\bibitem[Tegmark \& Bromley 1999]{tb99}
Tegmark, M., \& Bromley, B.C., 1999, ApJ, 518, L69

\bibitem[Tegmark \& Peebles 1998]{tp98}
Tegmark, M., \& Peebles, P.J.E., 1998, ApJ, 500, L79



\end{thebibliography}
\end{document}